A Heart for Interaction: Shared Physiological Dynamics and Behavioral Coordination in a
Collective, Creative Construction Task

Riccardo Fusaroli[a,b], Johanne S. Bjørndahl[b], Andreas Roepstorff[a] & Kristian Tylén[a,b]

A: The Interacting Minds Centre, Aarhus University, Jens Chr. Skous Vej 4, 8000 Aarhus,
Denmark

B: Center for Semiotics, Department for Aesthetics and Communication, Aarhus University, Jens
Chr. Skous Vej 2, 8000 Aarhus, Denmark

Corresponding author:

Riccardo Fusaroli

Address:

        Jens Chr. Skous Vej 2

        8000 Aarhus C

        Denmark

Phone:  (+45) 28890881

Email:  fusaroli@gmail.com

**Abstract**

Interpersonally shared physiological dynamics are increasingly argued to underlie rapport, empathy and even team performance. Inspired by the model of interpersonal synergy, we critically investigate the presence, temporal development, possible mechanisms and impact of shared interpersonal heart rate (HR) dynamics during individual and collective creative LEGO construction tasks. In Study 1 we show how shared HR dynamics are driven by a plurality of sources including task constraints and behavioral coordination. Generally, shared HR dynamics are more prevalent in individual trials (involving participants doing the same things) than in collective ones (involving participants taking turns and performing complementary actions). However, when contrasted against virtual pairs, collective trials display more stable shared HR dynamics suggesting that online social interaction plays an important role. Furthermore, in contrast to individual trials, shared HR dynamics are found to increase across collective trials. Study 2 investigates which aspects of social interaction might drive these effects. We show that shared HR dynamics are statistically predicted by interpersonal speech and building coordination. In Study 3, we explore the relation between HR dynamics, behavioral coordination, and self-reported measures of rapport and group competence. While behavioral coordination predicts rapport and group competence, shared HR dynamics do not. Although shared physiological dynamics were reliably observed in our study, our results warrant not to consider HR dynamics a general driving mechanism of social coordination. Behavioral coordination – on the other hand - seems to be more informative of both shared physiological dynamics and collective experience.

# 1. Introduction

As we go through our day, we continuously engage socially with other people: from chats at the bus stop, to joint projects at the work place, and nurture of long term relations with friends and family. Common to these otherwise very different social activities is that they depend, to varying degrees, on our ability to engage, align and closely coordinate actions and emotions with one another. Which processes are involved in the development of such coordination and rapport? Several studies have investigated aspects of complex coordinative dynamics - from mutual adaptations of movements, words and prosody (Christensen, Fusaroli, & Tylén, 2016; Fusaroli et al., 2012; Fusaroli & Tylén, in press; Riley, Richardson, Shockley, & Ramenzoni, 2011), to establishment of shared routines (Fusaroli, Raczaszek-Leonardi, & Tylén, 2014; Mills, 2014) – and their role in facilitating performance and rapport (Marsh, Richardson, & Schmidt, 2009). Moreover, a growing number of studies points to the crucial role of shared physiological dynamics: Interpersonal coupling has been observed at the level of brain activity (Dumas, Nadel, Soussignan, Martinerie, & Garnero, 2010; Friston & Frith, 2015; Hasson, Ghazanfar, Galantucci, Garrod, & Keysers, 2012; Konvalinka et al., 2014; Pickering & Garrod, 2013), but also low level physiological processes such as heart rate (henceforth HR) are found to become spontaneously coordinated among interacting individuals. Shared HR dynamics, in particular, are argued to underlie emotional common ground in social interactions, as they facilitate the construction of a sense of community (Konvalinka et al., 2011), empathy and mindreading (Levenson & Gottman, 1983; Levenson & Ruef, 1992), group social structure (Cleveland, Finez, Blascovich, & Ginther, 2012), and therefore might even underlie team performance (Elkins et al., 2009; Henning, Boucsein, & Gil, 2001; Henning & Korbelak, 2005; Strang, Funke, Russell, Dukes, & Middendorf, 2014; Wallot, Mitkidis, McGraw, & Roepstorff, submitted).

While these findings are indeed intriguing, no coherent framework has yet been developed to unravel the factors potentially leading to shared HR dynamics in the diverse contexts of social interactions. Critical investigation thus seems warranted. Are shared HR dynamics a property of mundane, everyday interactions? What are the mechanisms underlying them? And how are they connected to emotional common ground, rapport and joint performance?

A recent model – the *Interpersonal Synergies Model* as developed in (Fusaroli, Raczaszek-Leonardi, et al., 2014; Fusaroli & Tylén, in press) – provides a constructive starting point for an articulated investigation of shared HR dynamics in the context of social interactions. According to the Interpersonal Synergies Model, interpersonal coordination is contingent upon the affordances of the shared activity. That is, the behavior of interacting agents becomes interdependent (A's behavior impacts B's behavior, which in turn impacts A's behavior) *in ways that are shaped by their goals and activities*, and enabling the interacting agents to accomplish them. Analogous arguments are advanced by the enactive approach to social interactions (Cuffari, Di Paolo, & De Jaegher, 2014; Hanne De Jaegher & Di Paolo, 2008; H. De Jaegher, Di Paolo, & Gallagher, 2010; Froese & Di Paolo, 2010). In this approach, social coordination is conceived of as mutual enabling: coordination is not simple interdependence, but it is aimed at the successful accomplishment of the interactive activity. The experience of social interactions, thus, depends on how well coordination enables the success of the interaction itself (Froese, Iizuka, & Ikegami, 2014). As a consequence of these perspectives, any analysis of interpersonal coordination should start from an analysis of the demands of the shared activity as these constrain and guide the interaction. In a similar vein, the group's collective performance and experience of the activity will be related to how well the interpersonal coordination meets these demands.

Inspired by the synergy model, we devised a series of three studies to critically investigate shared HR dynamics in the context of a creative LEGO construction task. In the first study we investigate the impact of different task constraints: we assess the level and stability of shared HR dynamics contrasting individual and collective LEGO construction, thus contrasting the impact of simply doing the same activity (working individually side-by-side) and interacting (working together on a shared project). In the second study, we quantitatively assess the degree of interpersonal coordination of speech and construction activities involved in solving the collective tasks and their relation to shared HR dynamics. Finally, in the third study we assess the relation of these different forms of behavioral and physiological coordination on self-reported indices of the feeling of relatedness and collective competence. Together, the three studies allow for a better understanding of the conditions, mechanisms and impact of shared HR dynamics in the contexts of collaborative tasks, integrating this line of research with that on behavioral coordination.

In the following paragraphs we will situate our investigations in the context of current literature on shared HR dynamics to identify which mechanisms and impact have been found, before detailing our approach and findings.

## 1.1 Shared HR dynamics

A central aspect of human social coordination and collaboration concerns the way interacting individuals form group-level behaviors, for instance by co-regulating their emotional states. Heart rate seems a useful measure in this regard as it has been shown to increase with emotional arousal in ways that are emotion-specific (Ekman, Levenson, & Friesen, 1983; Levenson & Ruef, 1992; Schwartz, Weinberger, & Singer, 1981; Sinha, Lovallo, & Parsons, 1992). Heart rate is also connected with different aspects of behavior. Heart rate increases with physical activity (Wallot, Fusaroli, Tylén, & Jegindø, 2013) and it is (weakly) entrained to respiration rhythms (Schäfer, Rosenblum, Kurths, & Abel, 1998; Yasuma & Hayano, 2004). In other words, as interacting agents coordinate their behaviors and share emotional ground, they seem likely to also increasingly share HR dynamics.

Such intuitions have guided investigations of collective high emotional arousal in diverse social contexts: Di Mascio and colleagues (1955) showed shared co-variation between patient and psycho-therapist, and Levenson and Gottman (1983) observed high levels of shared physiological dynamics during couple therapy, although mostly in distressed couples during episodes involving negative emotions. Helm and colleagues (2012) found shared HR dynamics in romantic couples across a series of tasks aimed at eliciting shared emotional arousal. Müller and Lindenberger (2011) found similar patterns among choir singers, modulated by specific types of vocal coordination. Furthermore, Konvalinka and colleagues (2011) investigated shared HR dynamics in a high arousal context: a religious fire-walking ritual. Here shared HR dynamics were even found among participants doing quite different things: Individuals walking on burning coals and family members watching the ritual from the side. Interestingly, unrelated spectators did not share HR dynamics with firewalkers to the same extent.

A second line of research focused on more task-oriented collaborative settings, investigating the connection between shared HR dynamics and group performance. Henning and colleagues (Henning, Armstead, & Ferris, 2009; Henning, et al., 2001) showed that in collaborative maze-game and room-clearing tasks, teams displaying more shared HR dynamics solved the task faster and with fewer mistakes. As shared HR dynamics also correlated with self-reports of team coordination, task engagement, and performance (Gil & Henning, 2000), the researchers argued that they reflected increased team situation awareness, intra-team coordination, and shared mental models. To reinforce this point, Henning and Korbelak (2005) repeated the experiments introducing a disruption: at random times the team roles were re-organized. Immediately before such perturbations, shared HR dynamics positively predicted post-perturbation performance.

However, these results have recently been challenged. Strang and colleagues (2014) reported a negative correlation between shared HR dynamics, team performance and team attributes during a collaborative Tetris-like task. Analogously, Wallot and colleagues (Wallot, et al., submitted) reported shared HR dynamics to be negatively correlated with self-reports of collaborative satisfaction, and external raters' judgment of output product quality in a joint LEGO construction task.

Shared HR dynamics can thus be found in a variety of interpersonal contexts; however, the connection between heart rate, self-reported experience and collective performance seems far from a simple one. Findings seem conflicting and there is currently no consensus as to the factors regulating and impacting the level and dynamics of shared HR dynamics.

## 1.2 Contexts of shared HR dynamics

A first crucial issue concerns the mechanisms driving shared HR dynamics. It is clear that hearts (and autonomic systems in general) are not linked in any direct way. Nor are they easily intentionally coordinated. In other words, we need a better grasp on the mediating factors motivating and facilitating shared HR dynamics. The previously reviewed literature points to several, possibly interacting, factors: i) co-presence and history of previous interactions; ii) structure of the task that the interacting agents are sharing; and iii) degree of behavioral coordination between agents.

Simply being co-present in a situation has been observed to elicit behavioral and emotional contagion between individuals (Chartrand & Bargh, 1999; Hatfield & Cacioppo, 1994). People spontaneously mimic each other's behaviors and facial expressions, which can lead to the sharing of emotional states, with concomitant shared physiological arousal. Some studies of shared HR dynamics exploit this dimension of co-presence, focusing on how sharing the same situation induces shared physiological arousal in a way that is mediated by the history of previous interactions. For instance, being in a social, familiar or romantic relationship may catalyze shared HR dynamics (Chatel-Goldman, Congedo, Jutten, & Schwartz, 2014; Helm, et al., 2012; Konvalinka, et al., 2011; Levenson & Gottman, 1983).

However, in most studies people are not simply co-present, but share an activity. In these studies, people have to engage in a computer mediated tasks (Henning, et al., 2009; Henning, et al., 2001; Henning & Korbelak, 2005; Strang, et al., 2014), in the physical construction of a LEGO model (Wallot, et al., submitted), or in choir singing (Müller & Lindenberger, 2011). Each of these activities implies diverse constraints on the behaviors of the participants: sitting, being concentrated, breathing with a certain rhythm, etc. That fact that HR is influenced by physical activity and respiration might motivate the prediction that the task alone, by constraining participants' behaviors, drives important aspects of HR dynamics.

Finally, it has been pointed out that people entrain their behaviors in complex ways in contexts of social interactions: interacting individuals entrain their postural sways (Shockley, Santana, & Fowler, 2003), align to each other's behaviors and linguistic forms and complement each other's action (Dale, Fusaroli, Duran, & Richardson, 2013; Fusaroli, Raczaszek-Leonardi, et al., 2014; Fusaroli & Tylén, 2012, in press). Thus beyond mere co-presence and task constraints, online behavioral coordination might promote shared HR dynamics. The underlying mechanism of such interaction-driven HR dynamics could be behavioral synchronization (Louwerse, Dale, Bard, & Jeuniaux, 2012). Alternatively, shared arousal and coordination of complementary behaviors might be the dominating factors (Henning, et al., 2009; Konvalinka, et al., 2011; Strang, et al., 2014).

### 1.3 Impact of shared HR dynamics

A second open issue concerns what we can learn about human social relations, coordination, and collaboration from observations of shared HR dynamics. Henning and colleagues (Gil & Henning, 2000; Henning, et al., 2009; Henning, et al., 2001; Henning & Korbelak, 2005) argue that shared HR dynamics, by reflecting team members' sensorimotor and emotional integration, are indicative of the team's collaborative skills to the point that it can predict future performance. Other research, however, suggests caution as shared HR dynamics can also be driven by negative emotions (Levenson & Gottman, 1983), and dysfunctional team coordination and experiences (Strang, et al., 2014; Wallot, et al., submitted).

## 1.4 The structure of the investigation

To investigate these open issues, we designed an experimental study of HR and behavioral coordination in a creative construction activity. We manipulated aspects of the task constraints and assessed the impact on participants' coordination in order to investigate the functional nature of the coordinative dynamics. The experimental setup was inspired by the intervention method LEGO Serious Play (Gauntlett, 2007): Groups of participants repeatedly built LEGO models illustrating their understanding of six abstract notions. Sometimes they built the models collaboratively as a group, sometimes as individuals. We could thus compare the constraints of two different versions of the tasks. Individual constructions afforded high similarity of individual behaviors without online interaction: participants sat silently, concentrated and built each their individual model. Collective constructions, in contrast, afforded online social interaction and development of coordinative routines and strategies: group members could not all build or talk at the same time but had to coordinate their actions, take turns, negotiate joint understanding and resolve disagreements in order to complete the assignments (Bjørndahl, Fusaroli, Østergaard, & Tylén, 2014, 2015).

Relying on this setup, we devised three studies, to successively investigate the previously highlighted questions. In study 1, we focus on the way co-presence, task-constraints and online interaction impact shared HR dynamics, contrasting individual constructions requiring similar activities and collective constructions requiring actual coordination between group members in terms of complementary speech and building actions. In study 2, we ask about underlying mechanisms by investigating which aspects of behavioral coordination predict shared HR dynamics. In study 3 we explore the impact of shared physiological dynamics and behavioral coordination on self-reported experience of the interactions, in particular focusing on group members' experiences of relatedness and group performance.

## 2. Study 1 – Shared Heart Rate Dynamics

In study 1, we relied on a 2-by-2 factorial design with repeated measures to address the presence of and conditions for shared HR dynamics. Groups of 4-6 participants repeatedly constructed LEGO models individually and collectively. In both conditions we contrasted shared HR dynamics pairwise between participants co-present at the same table ('co-present' or 'real' pairs) and in pairs artificially constructed by selecting participants from two different groups, thus sharing the task, but not any actual co-presence or interaction ('separate' or 'virtual' pairs). The setup allows us to assess three hypotheses.

*Hypothesis 1: Co-presence*. If co-presence is crucial in facilitating shared HR dynamics, we will observe higher shared HR dynamics in real pairs than in virtual pairs, across individual and collective tasks.

*Hypothesis 2: Task-related affordances*. If task-related constraints are the driving force behind shared HR dynamics, we will observe higher shared HR dynamics in individual than in collective constructions, as the former involves participants simultaneously performing the same actions, while the latter often implies that participants to do different and complementary actions. However, real and virtual pairs should not display any significant difference, as it is task constraints and not co-presence or actual interaction driving shared HR dynamics.

*Hypothesis 3: Behavioral coordination*. If actual interpersonal behavioral coordination is driving shared HR dynamics, we will observe higher coordination in real pairs during collective tasks, but not in individual construction nor in virtual pairs. As behavioral coordination is likely to be established and develop over time, as participants develop turn-taking routines and strategies to co-build models (Bjørndahl, et al., 2015; Mills, 2014), the hypothesis also predicts increase in real pairs' shared HR dynamics over time in collective, but not in individual construction tasks. Notice that this hypothesis is further investigated in study 2.

It should be noted that these hypotheses are not fully independent, and that we expected to find multiple interacting sources of shared HR dynamics. In particular, the interpersonal synergy model strongly motivates hypotheses 2 and 3: Task constraints are a defining feature of the model and collective construction tasks crucially involve behavioral coordination.

## 2.1 Materials and Methods

### 2.1.1 Participants

30 participants (15 f, mean age 23.6, sd 2.6) were recruited among students of Aarhus University and received monetary compensation for their participation (ca. 50 $). All participants were native speakers of

Danish. Participants were organized in mixed-gender groups of four to six, randomly assembled. Group members did not know each other in advance. 1 participant was excluded due to a faulty device not recording heart rate data. This led to 5 groups with 5 participants and 1 with 4 (for a total of 29 participants).

### 2.1.2 Design and Procedure

Upon arrival, participants were strapped with heart rate monitor belts and wireless Lavalier microphones. Participants were placed around a table (fig 1) and familiarized with the materials and task through two practice trials. Two cameras video-recorded the group activities from different angles. HR sensors, microphones and cameras were all coupled to a central computer to allow precise temporal synchronization of the data.

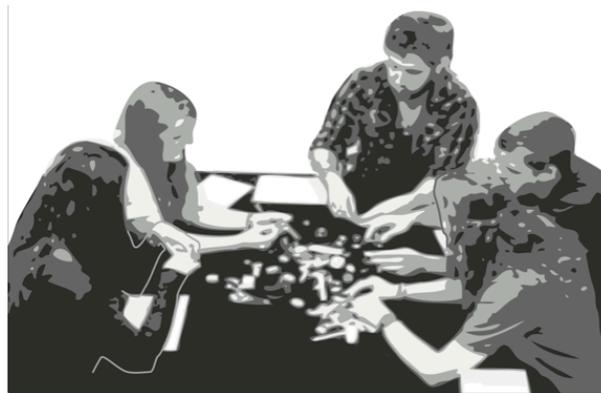

*Figure 1 – Experimental setup with five participants sitting around a table and collectively building a LEGO model*

The actual experiment was carried out as a two-condition within-group contrast: collective vs. individual. Each group underwent an interleaved series of six individual and six collective LEGO constructions of five minutes each, during which group members' heart rate was measured. In each trial, participants were instructed (in Danish) to use LEGO blocks to construct their understanding of one of six abstract concepts: 'responsibility' (ansvar), 'collaboration' (samarbejde), 'knowledge' (viden), 'justice' (retfærdighed), 'safety' (tryghed) and 'tolerance' (tolerance). The concepts were selected to be sufficiently common in public discourse that participants would know and have an opinion about them, but still challenging to construct in LEGOs. The LEGO materials consisted of a 'LEGO Serious Play Starter Kit' consisting of 214 mixed pieces (standards bricks in varying shapes and colors, wheels, LEGO people, etc.): one per participant in the individual and one per group in the collective construction tasks. In order to constrain variability of complexity and size of the models, participants were instructed to build their models within the limits of an A5 (5.8 x 8.3 inch) piece of cardboard (for an example of LEGO creations, see fig 2). An experimenter administered the instructions trial by trial, and the timing was carefully noted and double-checked in the video recordings.

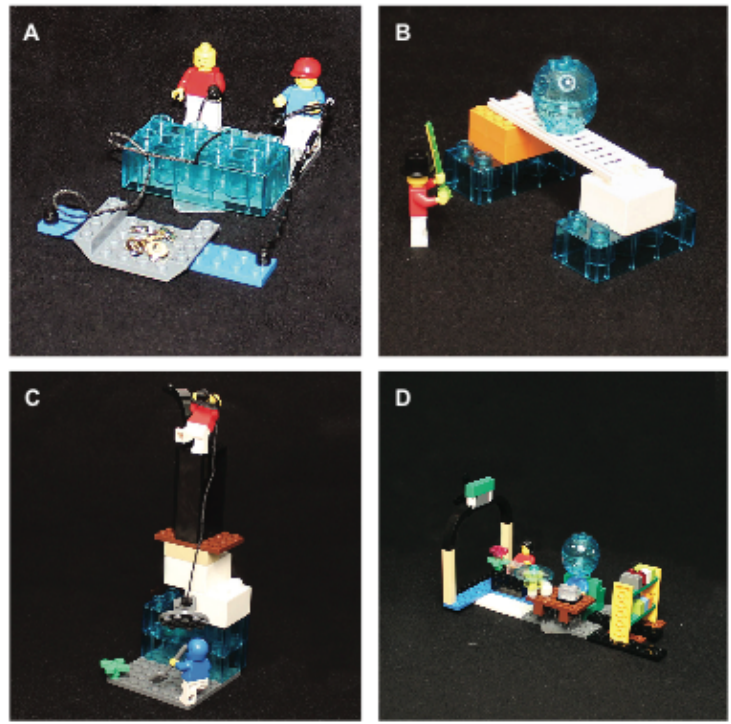

*Figure 2* –*Examples of LEGO models.* **A**: *Collective model illustrating "Collaboration".* **B**: *Collective model illustrating "Justice".* **C**: *Collective model illustrating "Responsibility".* **D**: *Collective model illustrating "Knowledge".*

Every concept was built twice, in succession, individually and collectively. The order was counterbalanced between trials. In individual trials, participants sat quietly and constructed their own models. In collective trials participants freely interacted to construct joint models. As the overall procedure took over 3 hours, the experiment was divided in two 3-concept sessions separated by a 20 min. break. Inspection of the videos confirmed that all participants were – although to various degree – engaged in the tasks (for details on qualitative aspects of the interactions during collective tasks, cf. Bjørndahl, et al., 2014, 2015).

### 2.1.3 Heart Rate Activity

Polar Team2 (Polar, 2013) chest-strapped heart rate monitors were used to record participants' heart beat activity as R-R intervals with millisecond accuracy. To align heart beat activity across participants we generated equally sampled time-series by estimating beats per minute every second based on sliding 5-second windows (Wallot, et al., 2013). We isolated 5-minute HR sequences corresponding to the 12 construction tasks.

### 2.1.4 Data Analysis: Interpersonal HR dynamics

HR time series present non-stationary dynamics. In other words, their means and standard deviations may vary over time as a function of e.g. activity, respiration and emotional arousal (Malik et al., 1996; Sayers, 1973). For this reason, linear methods assuming data stationarity, such as for instance correlation, are not appropriate to assess interpersonal shared HR dynamics. Additionally, interpersonal HR dynamics have been shown to display continuously varying lags of synchronization (Strang, et al., 2014), which require a method that does not assume constant lags as, for instance, a cross-correlation would do. We therefore chose to employ *Cross Recurrence Quantification Analysis* (CRQA), a method initially developed by Webber and Zbilut to quantify the shared dynamics of complex non-linear systems (Marwan, Thiel, & Nowaczyk, 2002; Shockley, Butwill, Zbilut, & Webber, 2002; Webber & Zbilut, 1994). Relying on two time-series, CRQA reconstructs the phase space of possible states and quantifies the structure of recurrence, that is, of the instances in which the two time-series display similar dynamics, controlling for individual baselines of HR (for more details on the methods, cf. Fusaroli, Konvalinka, & Wallot, 2014; Marwan, Carmen Romano, Thiel, & Kurths, 2007).

CRQA has been successfully employed to quantify the entrainment of coupled oscillators (Shockley, et al., 2002) and has been shown particularly effective in the analysis of interpersonal coordination, which often presents weak coupling of physiological (Konvalinka, et al., 2011; Strang, et al., 2014; Webber & Zbilut, 1994) and behavioral systems (Black, Riley, & McCord, 2007; Ramenzoni, Davis, Riley, Shockley,

& Baker, 2011; Richardson, Lopresti-Goodman, Mancini, Kay, & Schmidt, 2008; Shockley, Baker, Richardson, & Fowler, 2007; Shockley, et al., 2003). The use of recurrence techniques to assess HR dynamics has been validated and found more robust than alternative methods (Konvalinka, et al., 2011; Strang, et al., 2014; Wallot, et al., 2013).

With CRQA we were able to produce several metrics with which to estimate the similarity between dynamical patterns and capture many properties of the heart rate dynamics that would otherwise be lost due to averaging with more traditional correlation analysis. In particular, we calculated:

- *Level of coordination*, defined as the percentage of shared states in the phase space, that is the basic dynamic patterns that are repeated between the two time series (recurrence rate, RR). The higher the recurrence rate, the more similar the range of basic dynamic patterns displayed by the participants' HRs.

- *Stability of coordination*, defined as average length of sequences repeated across time-series (L). The higher the L, the more the participants tend to display prolonged and stable sequences of shared HR dynamics.

CRQA was calculated using the CRP Toolbox for Matlab 2014a. The phase space parameters were calculated according to Abarbanel (1996): the delay was set to minimize mutual information, and the embedding dimensions to minimize false nearest neighbors across all time-series. The threshold of recurrence was set to ensure recurrence rates comprised between 0.04 and 0.10 (Marwan et al 2007). This led to the following parameters: an embedding dimensions of 4, a delay of 6 and a threshold of 1.

### 2.1.5 Data Analysis: Virtual vs. Real Pairs

To assess the presence of interpersonal shared HR dynamics we calculated CRQA indexes for all possible pairs of members within each group (within-group real pairs) in both collective and individual construction tasks. Per each real pair we then randomly selected a participant and paired her with a participant from a different group, thus generating between-groups virtual pairs. This procedure generated 56 real pairs (10 pairs per each 5-participant group and 6 for the 4-participant group) and 56 virtual ones, measured during each of the 6 collective and 6 individual trials. Since we excluded HR time series in which the sensor lost contact for more than 5 seconds, we ended up with a total of 1306 data points (out of 1344).

**2.1.6 Data Analysis: Assessing the Presence and Temporal Development of shared HR Dynamics**

In order to assess Hypotheses 1 to 3, that is, compare the level and stability of shared HR dynamics according to the experimental manipulations, we employed mixed effects linear models, controlling for group and pair variability, as well as for order of conditions and possible session effects (as participants took a break midways through the experiment). Each index of shared HR dynamics (RR and L) was separately employed as dependent variable; activity (individual vs. collective), virtual vs. real, session and order of condition were used as fixed effects, pair and group were used as random effects including random slopes for activity and virtual vs. real. Given the limited amount of degrees of freedom in the data, we only looked at interactions between activity and virtual vs. real. Notice that the random effects structure constrains the degrees of freedom in the same way as a multi-level repeated measures model.

In order to assess the second part of Hypothesis 3, that is, the development of shared HR dynamics over time in the two experimental conditions, we again constructed a mixed effects linear model. Each index of shared HR dynamics (RR and L) was separately employed as dependent variable; activity, time (trial), session and order of condition were used as fixed effects, pair and group variability were used as random effects including random slopes for activity and time. Given the limited amount of degrees of freedom in the data, we only looked at interactions between activity and time. Post-hoc testing was performed to assess differences between individual and collective activity at time 1 and at time 6.

All models were tested for influential observations and normality of the residuals, and the assumptions were respected. Values reported and confidence intervals were based on bootstrapped analyses stratified by pair, group and tasks. Variance explained by the model was quantified using $R^2$: both marginal $R^2$, indicating the variance explained by fixed effect; and conditional $R^2$, indicating the variance of the full model (fixed and random effects) (Johnson, 2014). Per each predictor we reported unstandardized beta coefficient (ß), standard error of the coefficient, and p-value. Additionally, as hypothesis 2 relied on evidence for the null hypothesis, we calculated Bayes Factors (BF) with default priors for all of the predictors in all analyses. A Bayes Factor below 0.3 indicates substantial evidence for the null hypothesis, above 3 indicates substantial evidence for the alternative hypothesis, anything in between indicates anecdotal evidence at best (Morey, Rouder, & Jamil, 2014). Full details on the statistical results are reported in the tables, while the text only touches on the results relevant for the hypotheses investigated. The analyses were performed using the lme4, MuMIn, BayesFactor, boot and ggplot2 packages for R (3.1).

## 2.2 Results

### 2.2.1 Assessing Shared Physiological Dynamics

Detailed results of the analyses are reported in the tables: Table 1 for the level and Table 2 for the stability of shared HR dynamics. Hypothesis 1 (shared HR dynamics as an effect of co-presence) is not supported as level of shared HR dynamics is not significantly different for virtual and real pairs (cf. Table 1 and Figure 3a) and their stability is so only for collective and not individual construction tasks (cf. Table 2 and Figure 3b). Hypothesis 2 (shared dynamics as an effect of task affordances) is supported, though with an interesting twist. The level of shared HR dynamics fully supports the hypothesis: it is higher for individual construction tasks, with no difference between real and virtual pairs (cf. Table 1 and Figure 3a). The stability of shared HR dynamics, however, shows a different and more complex pattern. Virtual pairs display higher stability in individual than in collective tasks. However, real pairs show the opposite pattern: collective tasks induce higher stability than individual ones. Consequently, we also observe that real pairs display higher stability than virtual pairs during collective (real L: 4.801, CI: 4.678, 4.961; vs. virtual L: 3.638, CI: 3.518, 3.765; BF > 1000 ± 3.28%; a difference of ca. 4.8 seconds) but not individual tasks (real: 4.841, CI: 4.639, 5.069; vs. virtual: 4.660, CI: 4.497, 4.857; BF = 0.21 ± 5.3%), cf. Table 2 and Figure 3b. This difference supports hypothesis 3 (shared dynamics as an effect of actual coordination), as it suggests that actual interpersonal coordination affords higher stability of HR dynamics (cf. Table 2 and Figure 3b).

Table 1: Study 1 - Level of shared HR dynamics

| Variable | Factor | ß | SE | p-value | BF |
|---|---|---|---|---|---|
| RR | Real | 0 | 0 | 0.25 | 0.28 ±5.5% |
| $R^2$m=0.04 | Activity | -0.01 | 0 | 0.002* | >1000 |
| $R^2$=0.18 | | | | | ±3.96% |
| | Interaction | 0 | 0 | 0.91 | 0.08 ±4.01% |
| | Order | 0.01 | 0.01 | 0.04* | >1000 |

|  | | | | | ±7.14% |
| Session | 0.01 | 0 | 0.13 | 32.68 ±7.6% |

Table 2: Study 1 - Stability of shared HR dynamics

| Variable | Factor | ß | SE | p-value | BF |
| --- | --- | --- | --- | --- | --- |
| L | Real | 0.9 | 0.13 | <0.0001* | >1000 |
| $R^2$m=0.13 | | | | | ±2.32% |
| $R^2$=0.25 | Activity | -0.38 | 0.17 | 0.03* | 521.73 |
| | | | | | ±2.81% |
| | Interaction | 1.42 | 0.16 | <0.0001* | >1000 |
| | | | | | ±2.36% |
| | Order | 0.07 | 0.18 | 0.69 | 0.08±3.6% |
| | Session | 0.29 | 0.1 | 0.0049* | 8.79 |
| | | | | | ±5.82% |

Impact of activity on shared HR dynamics (real and virtual pairs)

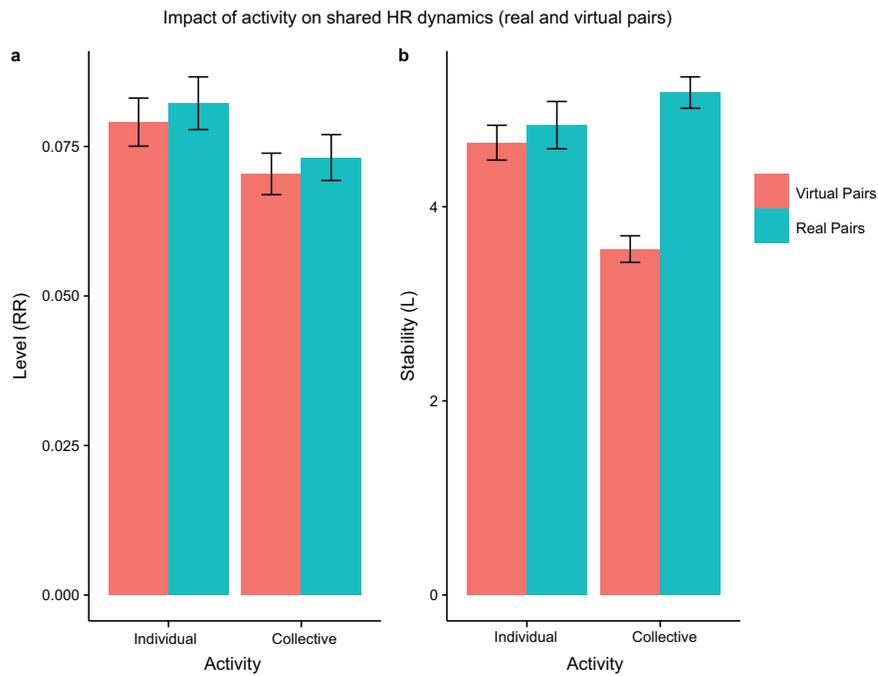

*Figure 3 – Study 1: Impact of activity on shared HR dynamics. Panel a on the left represents levels of shared HR dynamics individual and collective constructions and in real and virtual pairs. Panel b on the right analogously represents stability of shared HR dynamics.*

**Temporal Development of Coordination**

Detailed results of the analyses are reported in the tables: Table 3 for the level and Table 4 for the stability of shared HR dynamics. Hypothesis 3 is supported in that over time shared HR dynamics are shown to increase in collective but not individual tasks: both in level (collective: ß = 0.01, SE = 0.002, $p$ = 0.03, BF > 1000 ± 12.5%; individual ß = -0.01, SE = 0.005, $p$ = 0.015, BF = 1.52 ± 2.23%) and stability (collective ß = 0.25, SE = 0.11, $p$ = 0.018, BF > 1000 ± 1.96%; individual ß = -0.53, SE = 0.36, $p$ = 0.139, BF = 553.83 ± 3.47%), (cf. Table 3 and 4, and Figure 4). It is of particular interest that during the initial constructions shared HR dynamics are higher for individual than for collective constructions in both level (individual RR: 0.10, CI=0.09 0.11; collective RR: 0.07, CI = 0.07 0.08; BF > 1000 ± 3.11%) and stability (individual L: 5.66, CI: 5.24 6.11; collective L: 4.28, CI: 4.09 4.53, BF > 1000 ± 0.96%). This supports hypothesis 2. However, during the final constructions, temporal developments induce no difference in level of shared HR dynamics (individual RR: 0.07, CI=0.06 0.08; collective RR: 0.08, CI = 0.07 0.09, BF: 1.14 ± 3.02%) and higher stability during collective constructions when compared to individual ones (individual L: 4.71, CI: 4.39 5.11; collective L: 5.97, CI: 5.73 6.24, BF > 1000 ± 0.85%).

Table 3: Study 1 - Level of shared HR dynamics over time

| Variable | Factor | ß | SE | p-value | BF |
|---|---|---|---|---|---|
| RR | Activity | -0.01 | 0.006 | 0.0007* | 52.80 |
| $R^2$m=0.06 | | | | | ±1.31% |
| $R^2$=0.21 | Time | -0.01 | 0.003 | 0.001* | 4.16 |
| | | | | | ±1.73% |
| | Order | 0.01 | 0.006 | 0.07 | 125.08 |
| | | | | | ±3.38% |
| | Session | 0.02 | 0.006 | 0.002* | 68.43 |

| | | | | | ±4.01% |
|---|---|---|---|---|---|
| | Activity*Time | 0.004 | 0.001 | 0.0081* | 4.67 |
| | | | | | ±12.92% |

Table 4: Study 1 - Stability of shared HR dynamics over time

| Variable | Factor | ß | SE | p-value | BF |
|---|---|---|---|---|---|
| L | Activity | -0.89 | 0.36 | 0.003* | 2.70 |
| $R^2$m=0.04 | | | | | ±3.59% |
| $R^2$=0.18 | Time | -0.49 | 0.18 | 0.006* | 0.14 |
| | | | | | ±3.67% |
| | Order | 0.1 | 0.24 | 0.69 | 0.1 |
| | | | | | ±2.75% |
| | Session | 0.68 | 0.48 | 0.15 | 0.42 |
| | | | | | ±14.9% |
| | Activity*Time | 0.24 | 0.07 | 0.0008* | 29148.25 |
| | | | | | ±5.1% |

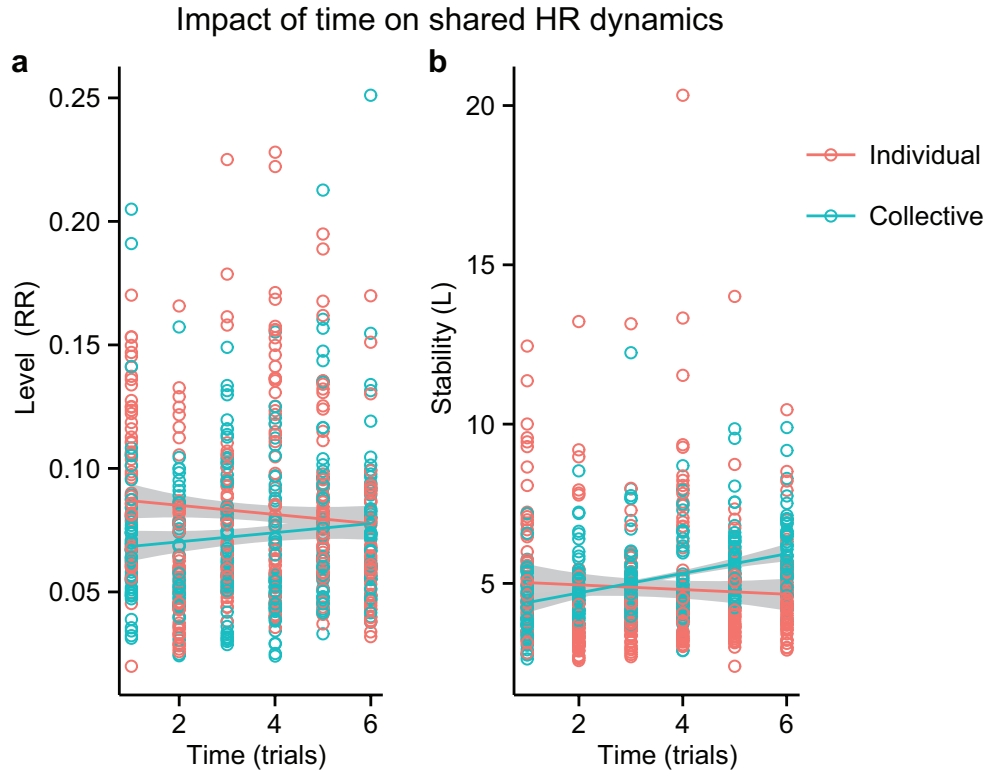

*Figure 4 – Study 1: Impact of time and activity on shared HR dynamics. Panel a on the left represents level*
*of shared dynamics and panel b on the right their stability.*

## 2.3 Discussion

The results support a combination of Hypothesis 2 and 3. Although we do not observe a general effect of co-presence (hypothesis 1) across individual and collective activity, neither in level nor in stability of shared HR dynamics, there is substantial evidence that task affordances (hypothesis 2) and online coordination (hypothesis 3) do play a role. The level of shared HR dynamics is largely driven by task constraints: engagement in the task makes participant entrain their heart rates irrespective of whether they are co-present or not as revealed by the fact that there was no significant difference between real and virtual pairs. Additionally, the significant difference between level of shared HR dynamics in individual and collective activity, points to different task constraints in these two conditions: while in individual constructions, participants do quite similar things (quietly concentrate on their individual LEGO model), in collective constructions, participants often structure their activities in more complementary ways (taking turns in talking and listening, holding the model, searching for and adding bricks, etc.). This variability in behavior across individuals in the collective trials is likely to lead to variability in HR, and thus lower levels of shared HR dynamics.

The stability of shared HR dynamics adds an interesting perspective to the story: while the overall level of shared HR dynamics is lower in collective trials, the stability is the same across collective and individual trials, for real pairs, but lower for virtual pairs. This suggests that actual online collaboration has an impact on shared HR dynamics above and beyond simple task constraints: collaborative tasks involve not just similar individual levels of engagement (being equally physiologically aroused), but repeated, prolonged sequences of behavioral coordination which might entrain physiological dynamics (see also Study 2). This hypothesis is further supported by the general increase in level and stability of shared HR dynamics over the course of collective but not individual trials; possibly as a function of coordinative routines being established. While the level of shared HR dynamics in collective constructions reaches individual constructions levels over the progression of trials, the stability even becomes significantly higher.

It should be noted that these effects are observed across groups. Groups are likely to develop different social structures according to emerging structures of dominance, interaction style, and reciprocal feedback (Bjørndahl, et al., 2015). This is expected to affect physiological states in the participants (Cleveland, et al., 2012). However, the fact that we can observe a general increase in shared HR dynamics above and beyond group differences suggests that the effect applies generally across diverse social structures. Future research will have to more explicitly investigate how social structure interacts with levels of physiological coordination and their temporal development.

### 3. Study 2: From behavioral coordination to shared physiological dynamics

Study 1 supported hypothesis 3 showing that certain properties of interpersonal shared HR dynamics, in particular its stability, are related to actual online interpersonal interaction during collective construction tasks. In order to further explore this hypothesis, we devised a second study more directly quantifying behavioral coordination and its relation to shared HR dynamics.

The collective construction of LEGO models requires at least two aspects of behavioral coordination: 1) participants have to discuss and agree upon their joint project, which requires taking turns in dialogical interaction; 2) participants cannot all manipulate the model at the same time but have to coordinate their actions in physically building the model. We have already discussed how motor activity impacts HR arousal (Wallot, et al., 2013). Speech activity is also likely to pose strong constraints on HR. Speaking has considerable effects on respiratory patterns, involving rapid inspiratory and prolonged expiratory phases (Hoit & Lohmeier, 2000; McFarland & Smith, 1989; Winkworth, Davis, Adams, & Ellis, 1995). More crucially, conversations constrain the temporal structure of individual speech contributions in turn taking.

Interlocutors involved in conversation have also been shown to spontaneously adapt to each other's linguistic forms, conversational moves, syntax and prosody (Fusaroli, Abney, Bahrami, Kello, & Tylén, 2013; Hopkins, Yuill, & Keller, 2015; Pickering & Garrod, 2004; Wilson & Wilson, 2005), which may lead to the development of interactional routines and procedural conventions (Fusaroli & Tylén, in press; Mills, 2014). By using more similar linguistic forms and coordinating the timing of their speech behavior, interlocutors might come to tightly interweave their breathing patterns (McFarland, 2001; Warner, Waggener, & Kronauer, 1983), which in turn have been shown to affect HR (Beda, Jandre, Phillips, Giannella-Neto, & Simpson, 2007; Berntson, Cacioppo, & Quigley, 1993).

Although no study, to our knowledge, has assessed the relative impact of the different behavioral modalities, it seems likely that the development of shared engagement, turn-taking and routines in speech and building activity will impact shared HR dynamics. In pursue of this hypothesis, Study 2 used the audio and video recordings of the Lego construction tasks to quantify interpersonal coordination of building and speech activity (BA and SA) and assessed its relation to shared HR dynamics.

## 3.1 Materials and Methods

### 3.1.1 Speech and Building Coding
In order to investigate the potential mechanisms influencing heart rate coordination, we quantified speech and building coordination during collective trials. Employing the coding software ELAN (Wittenburg, Brugman, Russel, Klassmann, & Sloetjes, 2006), two research assistants naive to the purpose of the study carefully screened the videos of the collective tasks and separately coded for presence (1) and absence (0) of speech and building activity for each participant on a second-by-second basis. The coding procedure thus provided us with a rough index of when and for how long individual participants were engaged in speech or building. Notice, however, that the way we capture the structure of engagement is insensitive to the specific speech and building behaviors at play. The coded data consisted of 5 groups of 5 and 1 group of 4 participants engaged in 6 collective trials (of five-minutes), which generated 170 speech engagement and 170 building engagement time series out of 180 total, the missing data being due to corrupted audio and video materials. Notice that the videos have also been analyzed qualitatively, with other research purposes in (Bjørndahl, et al., 2014, 2015) and show that participants were all engaged in the tasks.

**3.1.2 Speech and Building Coordination** Interpersonal coordination in speech and building engagement was quantified using CRQA on the nominal (1s and 0s) time series. While initially developed for numeric time-series CRQA has been widely employed for nominal time series (Dale, Warlaumont, & Richardson, 2011; Fusaroli & Tylén, in press; Louwerse, et al., 2012; Reuzel et al., 2013). Analogously to Study 1, CRQA allows us to quantify the level (RR) and stability (L) of shared behavioral engagement. Importantly, since CRQA assesses shared dynamics across multiple time-lags, it goes beyond simple synchrony and can thereby capture stable patterns of turn-taking. Parameters for CRQA were set according to Dale and colleagues (2010): embedding dimensions were set to 2, delay to 1 and threshold to 0. As we were interested in shared engagement, and not in recurrence of inactivity, we "blanked out" moments of inactivity by replacing 0's with -2000 in the first participant's time series and -3000 in the second participant's time series (thereby inactivity would not be counted as recurring between participants). We generated 56 real pairs assessed through 6 collective trials amounting to 320 data points (16 had to be excluded due to corrupted audio-video materials). Otherwise the procedure was analogous to Study 1.

**3.1.3 Data Analysis: Assessing the Impact of Behavioral Coordination on Shared Physiological Dynamics** In order to assess the relation between behavioral coordination and shared physiological dynamics, we employed mixed effects linear models with each of the indexes of heart rate coordination (RR and L) as dependent variable and all four indexes of speech (Speech RR and L) and building (Build RR and L) coordination as fixed effects. We used only random intercepts for group and pair variability, as the models with random slopes did not converge. The model employed 310 data points, as we had to exclude data points where either shared HR dynamics or behavioral coordination values were not present. The procedure was otherwise analogous to Study 1.

**3.2 Results**

The level of heart rate coordination (HR RR) was significantly related to the level of building and the stability of speech coordination (Build RR and Speech L): the higher the level of building coordination and the more stable the speech coordination, the higher the level of shared HR dynamics, with particularly substantial evidence in favor of the role of stable coordination (cf. Table 5 and Figure 5). The stability of heart rate coordination (HR L) was significantly and with substantial evidence related to speech coordination: the more frequent (RR) and stable (L) the speech coordination, the more stable the heart rate entrainment  (cf. Table 5 and Figure 5).

Table 5: Study 2 - From Behavioral Coordination to Shared Physiological Dynamics

| Variable | Factor | ß | SE | p-value | BF |
|---|---|---|---|---|---|
| HR RR | Build RR | 0.23 | 0.12 | 0.045* | 0.96 |
| $R^2$m=0.14 | | | | | ±22.89% |
| $R^2$=0.23 | Build L | ≈0 | ≈0 | 0.248 | 0.32 |
| | | | | | ±23.4% |
| | Speech RR | 0.01 | 0.01 | 0.456 | 0.53 |
| | | | | | ±22.45% |
| | Speech L | ≈0 | ≈0 | 0.0013* | 4.67 |
| | | | | | ±18.66% |
| HR L | Build RR | 2.04 | 4.19 | 0.63 | 0.18 |
| $R^2$m=0.37 | | | | | ±14.17% |
| $R^2$=0.47 | Build L | -0.16 | 0.09 | 0.077 | 0.63 |
| | | | | | ±19.86% |
| | Speech RR | 1.77 | 0. 45 | <0.0001* | >1000 |
| | | | | | ±22.69% |
| | Speech L | 0.02 | ≈0 | <0.0001* | 194.33 |
| | | | | | ±18.55% |

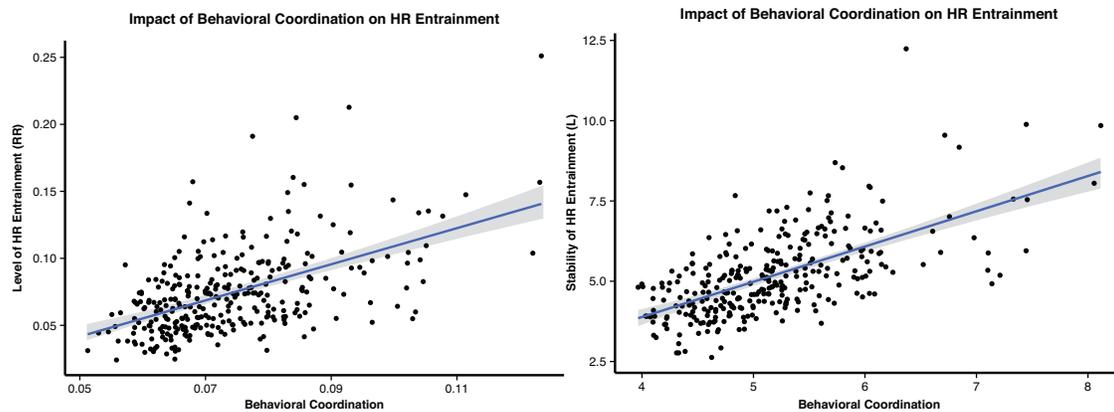

*Figure 5: Study 2: From Behavioral Coordination to Shared Physiological Dynamics: In the first panel on the left we represent the model predictions against the level of shared HR dynamics scores. In the second panel on the right we represent the model predictions against the stability of shared HR dynamics scores.*

### 3.3 Discussion

In Study 1, we observed that the stability of shared HR dynamics in collective activity was higher in real than in virtual pairs suggesting a role for actual online behavioral coordination (hypothesis 3). Study 2 further supports this hypothesis: the level and stability of speech and building coordination positively predict shared HR dynamics, suggesting that online behavioral coordination is actively shaping shared HR dynamics. While the level of shared HR dynamics is weakly, but significantly predicted (2% of the variance explained), its stability is more strongly predicted (25% of the variance explained). The results point to an intriguing complexity of mechanisms: while the level of physiological entrainment is (weakly) related to motor and speech coordination, its stability is connected to speech coordination alone. Building and speech coordination thus seem to contribute to shared HR dynamics in complementary ways.

## 4. Study 3 – Coordination, Shared Physiological Dynamics and Experience

While study 2 supported the hypothesis that shared HR dynamics are – at least partially – driven by behavioral coordination, another question regards how participants' subjective experience of the collaborative interactions relates to behavioral coordination and shared HR dynamics. In previous studies, behavioral coordination and shared HR dynamics have each been separately related to the phenomenological experience of social interactions (Gil & Henning, 2000; Marsh, et al., 2009). However, as shared HR dynamics are likely to be mediated through behavioral coordination, indices of behavioral coordination might outperform the physiological entrainment in predicting experiential variables. In Study 3 we thus compared the predictive power of behavioral coordination and physiological entrainment on self-rated experience. In particular, we focused on the participants' experiences of relatedness to their fellow group members and assessment of their group's competence in solving the task.

## 4.1 Materials and Methods

**4.1.1 Experience of the Collaboration** In order to provide insight into group members' experience of the creative collaboration, at the end of the experiment all participants filled in the Intrinsic Motivation Inventory (IMI, Ryan, 1982). In the following we will especially consider the two factors *Relatedness to the group* and *Perceived collective competence* since they relate most directly to the quality of interpersonal coordination. Notice that the creative LEGO construction tasks did not afford objective analysis of performance (e.g. "quality" of resulting models). However, general correlations between group cohesion, self-reported competence and externally measured performance have often been reported (Lyons, Funke, Nelson, & Knott, 2011; Paskevich, Brawley, Dorsch, & Widmeyer, 1995).

**4.1.2 Data Analysis: From Coordination to Experience** As experience was measured at the level of individual participants, but referred to their experience of the group, we calculated a coordination index for each individual group member by averaging all within-group pair scores to which an individual participated. In order to assess the potential impact of behavioral and physiological coordination on experience, we employed mixed effects linear models with the 2 indexes of experience separately as dependent variables, and the individual indexes of interpersonal coordination as fixed factors, while group was a random factor. As experience was assessed at the end of the experiment, we chose to employ indexes of interpersonal coordination from the last trial of the experiment. The models involved 23 data points, as the audio-video recordings of a group's last task was corrupted, one participant had a malfunctioning HR sensor, and one participant did not complete the questionnaire.

## 4.2 Results

Self reported relatedness was significantly predicted by the stability of building and speech coordination: the more stable building and the less stable speech coordination, the higher the ratings. Self reported group competence was significantly predicted by the level of building coordination: the less building coordination, the higher the ratings (cf. Tables 6 and 7 and Figure 6). None of the HR predictors were significantly correlated with experience.

Table 6: Study 3 - Relatedness and Interpersonal Coordination

|  | ß | SE | p-value | BF |
|---|---|---|---|---|
| $R^2$m: 0.45 | | | | |
| $R^2$:   0.45 | | | | |
| HR RR | -69.47 | 122.03 | 0.57 | 0.7 ±2.45% |
| HR L | 2.69 | 3.61 | 0.46 | 0.75 ±2.45% |
| BA RR | -133.05 | 151.02 | 0.38 | 0.65 ±2.42% |
| BA L | 7.80 | 3.59 | 0.03* | 1.51 ±2.52% |
| SA RR | 4.64 | 13.17 | 0.73 | 0.66 ±2.45% |
| SA L | -0.28 | 0.14 | 0.046* | 0.82 ±2.42% |

Table 7: Study 3 - Group Competence and Interpersonal Coordination

|  | ß | SE | p-value | BF |
|---|---|---|---|---|
| $R^2$m: 0.27 | | | | |
| $R^2$:   0.38 | | | | |
| HR RR | 0.93 | 69.38 | 0.18 | 1.34 ±2.4% |
| HR L | -1.67 | 2.04 | 0.41 | 0.85 ±2.32% |
| BA RR | -0.02 | 82.12 | 0.01* | 3.53 ±2.72% |
| BA L | 3.08 | 2.03 | 0.13 | 1.66 ±2.5% |
| SA RR | 3.08 | 7.8 | 0.69 | 0.73 ±2.31% |
| SA L | 0 | 0.09 | 0.93 | 0.7 ±2.28% |

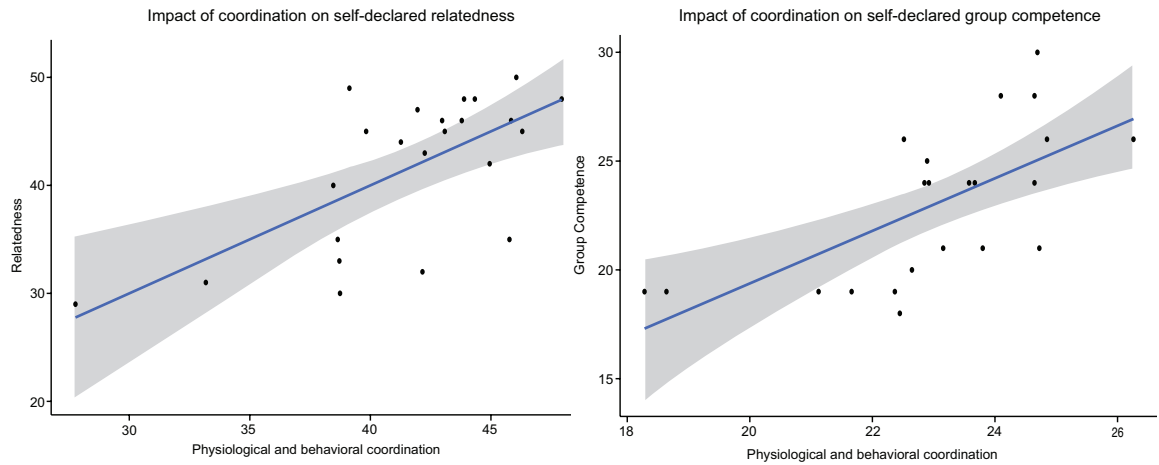

*Figure 6: Study 3: From Interpersonal Coordination to Experience: In the first panel on the left we represent the model predictions against the Relatedness scores. In the second panel on the right we represent the model predictions against the Group Competence scores.*

## 4.3 Discussion

The results suggest that behavioral coordination (building and speech) is a better predictor of the phenomenological experience of group relatedness and competence than shared HR dynamics. This supports the idea that emotions might be environmentally extended, and more specifically dependent on specific sustained patterns of social interactions (Colombetti & Krueger, 2014; Krueger, 2014). Curiously, HR does not correlate with experience at all. This suggests that in collaborative tasks – in which behavioral coordination is crucial – behavioral coordination mediates both physiological entrainment and experience. However, it should be noted that the connection between behavioral coordination and experience is not a trivial one: some of the indexes of interpersonal behavioral coordination are negatively related to experience. This negative correlation between coordination and experience has been previously observed in (Strang, et al., 2014; Wallot, et al., submitted) and might point to the complexity of coordination dynamics and their timescales. Good cooperation cannot always be reduced to doing the same things or even continuously coordinating, and division of labor might play an important role (Bjørndahl, et al., 2015; Fusaroli, Raczaszek-Leonardi, et al., 2014). Future studies will have to look into the more fine-grained details of speech and motor behaviors to better understand the nature of the coordination being established.

# 5. General Discussion

Our findings present a nuanced perspective on the complex nature of shared HR dynamics and their role in social interactions. Even in a relatively low arousal task, such as LEGO construction tasks, we find reliable modulations of shared HR dynamics between collaborating individuals. A number of factors seem to impact the level and stability of shared HR dynamics. Shared HR dynamics can be shown to vary according to the demands of the task: generally, individual trials induce higher level of shared HR dynamics than collective trials. We speculate that this is because in individual tasks, participants engage in very similar actions (quietly build each their model). However, only collective trials show the marks of actual interactions: within-group pairs display a significantly higher stability of shared HR dynamics than virtual pairs in collective, but not in individual trials. Similarly, shared HR dynamics grows during collective, but not during individual trials. Further supporting this line of arguments, shared HR dynamics during collective trials were predicted by level and stability of speech and building coordination: the more behavioral coordination, the more shared HR dynamics. Consistently, we observed behavioral coordination (but not shared HR dynamics) to predict perceived group relatedness and competence.

Our findings both articulate and challenge previous findings in the literature on shared HR dynamics. They corroborate the hypothesis that collective settings involve shared HR dynamics, and articulate two separate aspects of them – level and stability -, but they also question their potential role as foundational mechanism for shared experience, rapport and collective performance.

Shared HR dynamics can be argued to rely on several interacting factors. For instance, the structure of the task – e.g. having to build a LEGO model – seems to drive the level of shared HR dynamics as it affects and constrains the general amount of engagement and physiological arousal in participating individuals. However, the fine-grained dynamics of shared physiological activity seems to be related to the actual unfolding and development of interpersonal behavioral coordination. Indeed, behavioral coordination is a weak predictor of the level of shared HR dynamics (Building RR and Speech L) and a strong predictor of its stability (Speech RR and L). The collective LEGO construction task requires participants to develop coordinative strategies at many levels: negotiating concepts, and taking turns in speaking and building the models (Bjørndahl, et al., 2014, 2015). By developing shared behavioral routines, jointly regulating the group's action and speech (and thus indirectly respiration), the participants become partially entrained even at a physiological level. Interestingly, these behavioral routines, with their complementary roles (A speaks, B listens; A holds the model, B adds a brick) are much more informative about the emotional environment and competence of the groups than the physiological activity. In other words, shared HR dynamics do not seem to play a causal role: participants do not need to synchronize their hearts to feel related or coordinate their speech and building actions. However, they might need to coordinate their behavior in systemic ways in order to effectively solve the task at hand, and that in turn is reflected in the shared dynamics of their physiological states.

Our findings inform general discussions of the underlying mechanisms of human social coordination. Many studies have pointed to low level automatic priming mechanisms as responsible for many aspects of human social behavior (Garrod & Pickering, 2004; Pickering & Garrod, 2004; Richardson, Marsh, Isenhower, Goodman, & Schmidt, 2007; Shockley, et al., 2003). While such accounts capture crucial aspects of interpersonal adaptation and entrainment, they often do not take into consideration i) the fundamentally complementary nature of human coordination, and ii) the active, shaping role of the functional context - such as the structure of the joint task - on social coordination. Taking as point of departure the Interpersonal Synergies Model and the enactive approach to social interactions (Froese, Paolo, & Ezequiel, 2011; Fusaroli, Raczaszek-Leonardi, et al., 2014; Fusaroli & Tylén, in press), has allowed us to critically scrutinize the extent to which human behavioral and physiological coordination arise in response to a number of interacting factors pertaining to task constraints, the temporal stabilization of shared routines, and aspects of individual experience. Our findings warrant caution when interpreting observations of social coordination:  shared behavioral and physiological dynamics are not a universal panacea, automatically creating shared emotions, empathy and performance. Rather, the particular structure of the task seems to be a crucial factor in determining which aspects of interpersonal coordination selectively facilitate the development of rapport and collective performance. Future research should contrast tasks with different coordinative affordances and groups with different social structures to investigate additional sources of shared physiological dynamics and more precisely pinpoint the physiological mechanisms through which behavioral coordination impacts shared physiological dynamics, and their relation to experience and performance.

Finally it should be investigated whether our findings are specific to HR dynamics or could be more generally extended to other peripheral physiological dynamics such as electrodermal activity and blood pressure. We have no special reason to hypothesize radically different dynamics in different physiological measures. The reaction times of the physiological mechanisms, however, might vary and therefore affect the possibilities and time scales of physiological coordination. More research is needed to better map the relative dynamics of different peripheral physiological measures and their relation to behavior, emotions and coordination (Cacioppo & Tassinary, 1990; Young & Benton, 2014).

## 6. Conclusion

The findings in our study suggest that shared physiological dynamics can be found in collective activities and are influenced by a plurality of factors such as structure of the task and the behavioral coordination among interacting individuals. They also point to the fact that shared physiological dynamics should not be considered the unmediated proxy for shared emotions, empathy and collective performance. Rather behavioral coordination – at least in tasks requiring forms of joint action – seems to be driving both interpersonal physiological dynamics and collective experience.

## Acknowledgments

We would like to thank Ivana Konvalinka, Linda Post, John McGraw, Robert Rasmussen, Bo Stjerne Thomsen, and the LEGO Foundation. The research in this paper was funded by the Danish Council for Independent Research – Humanities grant "Joint Diagrammatical Reasoning"; the ESF EuroUnderstanding grant "Digging for the Roots of Understanding" and the Interacting Minds Centre, Aarhus University.

## Bibliographical References